\documentclass[twocolumn, trackchanges]{aastex62} %{amsart}

\usepackage{xspace}
\usepackage{color}
\usepackage{rotating}
\usepackage{comment}
\usepackage{amsmath}
\usepackage{subfigure}
\usepackage[ampersand]{easylist}
\usepackage[version=4]{mhchem} % Molecular species! e.g. \ce{H2O}
\usepackage[normalem]{ulem}    % for strikeout text use: \sout{removed text}
\usepackage[nameinlink]{cleveref}
\usepackage{upgreek}

\PassOptionsToPackage{hyphens}{url}\usepackage{hyperref}

\newcommand{\smarter}{\texttt{smarter}\xspace}
\newcommand{\smart}{\texttt{smart}\xspace}
\newcommand{\lblabc}{\texttt{lblabc}\xspace}
\newcommand{\dynesty}{\texttt{dynesty}\xspace}

\newcommand{\pandexo}{\texttt{PandExo}\xspace}

\def\gsim{~\rlap{$>$}{\lower 1.0ex\hbox{$\sim$}}}
\def\lsim{~\rlap{$<$}{\lower 1.0ex\hbox{$\sim$}}}
\newcommand{\emcee}[0]{\texttt{emcee}\xspace}

\newcommand{\um}[0]{$\upmu$m\xspace}

\begin{document}

%% ApJ style
\bibliographystyle{apj}

%% Slugcomment
%\slugcomment{Draft version \today}

\title{Hierarchical Bayesian Atmospheric Retrieval Modeling for Population Studies of Exoplanet Atmospheres: \\A Case Study on the Habitable Zone}

%% Short title, authors
\shorttitle{HBAR for the HZ}
\shortauthors{Lustig-Yaeger et al.}

\correspondingauthor{Jacob Lustig-Yaeger}
\email{Jacob.Lustig-Yaeger@jhuapl.edu}

\author[0000-0002-0746-1980]{Jacob Lustig-Yaeger}
\affiliation{Johns Hopkins University Applied Physics Laboratory, Laurel, MD 20723, USA}
\affiliation{NASA NExSS Virtual Planetary Laboratory, Box 351580, University of Washington, Seattle, Washington 98195, USA}

\author[0000-0001-7393-2368]{Kristin S. Sotzen}
\affiliation{Johns Hopkins University Applied Physics Laboratory, Laurel, MD 20723, USA}
\affiliation{Johns Hopkins University, 3400 N. Charles Street, Baltimore, MD 21218, USA}

\author[0000-0002-7352-7941]{Kevin B. Stevenson}
\affiliation{Johns Hopkins University Applied Physics Laboratory, Laurel, MD 20723, USA}

\author[0000-0002-0296-3826]{Rodrigo Luger}
\affiliation{Center~for~Computational~Astrophysics,~Flatiron~Institute,~New~York,~NY}
\affiliation{NASA NExSS Virtual Planetary Laboratory, Box 351580, University of Washington, Seattle, Washington 98195, USA}

\author[0000-0002-2739-1465]{Erin M. May}

\author[0000-0002-4321-4581]{L. C. Mayorga}
%\affiliation{Johns Hopkins University Applied Physics Laboratory, Laurel, MD 20723, USA}

\author[0000-0001-8397-3315]{Kathleen Mandt} 
%\affiliation{Johns Hopkins University Applied Physics Laboratory, Laurel, MD 20723, USA}

\author[0000-0003-1629-6478]{Noam R. Izenberg} 
\affiliation{Johns Hopkins University Applied Physics Laboratory, Laurel, MD 20723, USA}

%% Abstract %%

\begin{abstract}

With the growing number of spectroscopic observations and observational platforms capable of exoplanet atmospheric characterization, there is a growing need for analysis techniques that can distill information about a large population of exoplanets into a coherent picture of atmospheric trends expressed within the statistical sample. 
In this work, we develop a Hierarchical Bayesian Atmospheric Retrieval (HBAR) model to infer population-level trends in exoplanet atmospheric characteristics. 
We demonstrate HBAR on the case of inferring a trend in atmospheric \ce{CO2} with incident stellar flux, predicted by the presence of a functioning carbonate-silicate weathering negative feedback cycle, an assumption upon which all calculations of the habitable zone (HZ) rest. 
Using simulated transmission spectra and JWST-quality observations of rocky planets with \ce{H2O}, \ce{CO2}, and \ce{N2} bearing atmospheres, we find that the predicted trend in \ce{CO2} causes subtle differences in the spectra of order 10 ppm in the $1-5$ \um range, underscoring the challenge inherent to testing this hypothesis. In the limit of highly precise data (100 stacked transits per planet), we show that our HBAR model is capable of inferring the population-level parameters that characterize the trend in \ce{CO2}, and we demonstrate that the null hypothesis and other simpler trends can be rejected at high confidence. 
Although we find that this specific empirical test of the HZ may be prohibitively challenging in the JWST era, the HBAR framework developed in this work may find a more immediate usage for the analysis of gas giant spectra observed with JWST, Ariel, and other upcoming missions.   

\end{abstract}

%%% Keywords %%%

\section{Introduction} 
\label{sec:intro}

The habitable zone (HZ) provides a tangible starting point in the search for habitable exoplanet surface environments and life beyond the solar system \citep{Kasting1993, Kopparapu2013, Kaltenegger2017, Meadows2018c}.  
Despite the HZ's scientific lineage rooted in Earth system science \citep[e.g.,][]{Hart1978, Hart1979}, understanding the persistent habitability of Earth over geologic time remains an ongoing interdisciplinary investigation \citep[e.g.,][]{Goldblatt2011, Charnay2020, Isson2020, Stueken2020}, even before the principles of Earth's habitability are extended into the lesser understood exoplanet sample. Rather than undercutting the usefulness of the HZ, these underlying Earth-centric assumptions form the basis of a compelling hypothesis on the general nature of planetary habitability that has yet to be observationally tested using exoplanets, and may one day feed back into our understanding of Earth \citep{Shorttle2021, Komacek2021}.   

This connection is well exemplified by the carbonate-silicate weathering negative feedback cycle \citep{Walker1981}, which is thought to have helped maintain habitable surface conditions on Earth over billions of years via atmospheric \ce{CO2} buffering \citep{Berner2003}, but which is also an assumed ingredient in habitable zone calculations \citep{Kasting1993, Williams1997, Kopparapu2013}. In the carbonate-silicate weathering cycle, atmospheric \ce{CO2} warms the climate via the greenhouse effect. An increase in volcanic outgassing produces more \ce{CO2}, which further warms the climate. However, the weathering rate of continents also increases with temperature, thereby increasing the rate at which \ce{CO2} is removed from the atmosphere and ultimately subducted back into the mantel. Thus, the temperature dependence of the weathering rate provides the climate-stabilizing negative feedback that helps to maintain habitable surface temperatures against increases in \ce{CO2} and the solar luminosity over geologic timescales \citep{Glaser2020}. 

The link between Earth's long-term climate evolution and our perspective on exoplanet habitability provides a compelling opportunity to observationally test such hypotheses on the nature of planetary habitability using the population of exoplanets. 
\citet{Bean2017} outlined a statistical comparative planetology approach to empirically test the habitable zone hypothesis by recognizing that the climate model calculations for the HZ form a set of predictions that can be tested in the future using the growing sample of known likely-rocky exoplanets. Specifically, if the carbonate-silicate weathering feedback mechanism operates roughly as expected by climate theory, \textit{rocky exoplanets should exhibit a trend of increasing atmospheric \ce{CO2} from the inner edge of the HZ to the outer edge such that temperate surface temperatures are maintained.} 
\citet{Lehmer2020} used a coupled climate and weathering model to investigate the dependence of this trend on practical geophysical and physiochemical differences that are likely to exist between exoplanets. They reaffirmed the existence of such a trend in their models, but found that significant scatter may make it difficult to distinguish observationally, instead suggesting that the 2D distribution of planets in the flux-\ce{CO2} phase space may yield a more reliable test of the HZ hypothesis.  
Thus, although no single exoplanet can offer a definitive test of the habitable zone, the entire exoplanet ensemble provides a population that, in theory, can. 

However, testing for a statistical comparative planetology trend in atmospheric composition will be challenging because the trend itself is not directly observable, but must be properly identified by synthesizing the results of many individual inferences. Since exoplanet atmospheric compositions must be inferred from spectroscopic observations using retrieval models, any trends in atmospheric composition fall into the category of multilevel or hierarchical inference problems. While numerous trends in exoplanet atmospheric composition have been suggested as a means to understand exoplanet habitability \citep[e.g.,][]{Turbet2019, Checlair2019, Bixel2020, Checlair2021, Bixel2021}, a consistent framework to tackle the hierarchical atmospheric retrieval problem---going from spectroscopic observations to population-level atmospheric trends---has not been presented. 

In this paper, we present a novel retrieval methodology that enables inferences from observations of multiple planets to be combined and synthesized to constrain population-level atmospheric characteristics, and we apply the model to an idealized population of potentially habitable exoplanets to test for the predicted carbonate-silicate weathering \ce{CO2} trend. This is achieved using a new hierarchical Bayesian atmospheric retrieval (HBAR) modeling approach using the importance sampling formalism from \citet{Hogg2010}. This use case is a natural extension of a classical hierarchical Bayesian parameter estimation problem \citep{Gelman2013}, which have been highly successful in numerous exoplanet population studies \citep[e.g.][]{Hogg2010, Rogers2015, Wolfgang2016}, and recently applied to the atmospheric characterization of hot Jupiters using \textit{Spitzer} eclipse measurements \citep{Keating2021}. However, such methods have yet to be implemented for exoplanet atmospheric retrievals. Typically, hierarchical models are used to properly account for and determine the underlying population-level prior distributions from which an entire population of astrophysical objects are sampled. For instance, the mass-radius relationship has been constrained for sub-Neptune sized planets using mass and radius inferences across an ensemble of known exoplanets \citep{Wolfgang2016}. While completely novel to exoplanet atmospheric retrievals, a similar approach has been applied to Earth remote sensing aerosol retrievals by leveraging a hierarchical model with a built-in spatial dependence to capture spatial smoothness \citep{Wang2011}. 

In Section \ref{sec:methods}, we describe our standard and hierarchical retrieval methods. In Section \ref{sec:results}, we present the results of our atmospheric retrieval modeling, and then use them to infer \ce{CO2} trends with our HBAR model and perform a population-level model comparison. In Section \ref{sec:discussion} and Section \ref{sec:conclusion} we provide a discussion and conclusion of our findings, respectively. 

\section{Methods} 
\label{sec:methods} 

We begin by describing our nominal exoplanet atmospheric retrieval model in Section \ref{sec:methods:nominal}, which shares many common traits with other retrieval codes in the literature. We then detail the hierarchical modeling approach and how it interfaces with the standard retrieval framework in Section \ref{sec:methods:hbar}. 

\subsection{Nominal Retrieval Model}
\label{sec:methods:nominal}

We use the Spectral Mapping Atmospheric Radiative Transfer for Exoplanet Retrieval model (\smarter) to solve the Bayesian inverse problem on simulated terrestrial exoplanet transmission spectra (\citealp{Lustig-Yaeger2020thesis}; Lustig-Yaeger et al. 2021, in prep). We provide a brief description of the \smarter model below, which is limited to the essential components for this work, but refer the reader to \citet{Lustig-Yaeger2020thesis} and Lustig-Yaeger et al. (2021, in prep) for a complete description of the \smarter retrieval model and its rigorous validation using exoplanet-analog observations of Earth's infrared transmission spectrum. 

\subsubsection{Forward Model}
\label{sec:methods:forward}

\smarter relies on the Spectral Mapping Atmospheric Radiative Transfer model (\smart) as the core of the forward model used to simulate line-by-line transmission spectra for transiting exoplanets \citep{Meadows1996, Crisp1997, Misra2014a} using the ray tracing formalism described in \citet{Robinson2017a}. In turn, \smart leverages the DIScrete Ordinate Radiative Transfer \citep[DISORT;][]{Stamnes2017DISORT} model to solve the radiative transfer equation. 
The Line-By-Line ABsorption Coefficient code \citep[\lblabc; developed by D. Crisp;][]{Meadows1996} is used to calculate molecular vibrational-rotational absorption coefficients for input into \smart radiative transfer calculations. \lblabc combines information about the atmospheric state with HITRAN line-parameter and isotope information from the HITRAN2016 line list \citep{Gordon2017} to calculate gas absorption coefficients as a function of pressure, temperature, and wavenumber. Collisionally-induced absorption (CIA) data are used for \ce{CO2-CO2} \citep{Moore1972, Kasting1984, Gruszka1997, Baranov2004, Wordsworth2010, Lee2016} and \ce{N2-N2} \citep{Lafferty1996, Schwieterman2015b}.  

For simplicity in this study we consider one-dimensional atmospheres composed of \ce{N2}, \ce{CO2}, and \ce{H2O} with isothermal temperature-pressure (TP) profiles and evenly-mixed gas abundances. While plainly limited, this combination of gases is consistent with the climate modeling work that underpins the HZ \citep[e.g.][]{Kasting1993, Kopparapu2013}. We allow the ($\log_{10}$) volume mixing ratios (VMRs) of \ce{CO2} and \ce{H2O} to freely vary within the forward model, but set the \ce{N2} VMR to the residual VMR such that the VMRs of all three gases sum to unity, as in numerous retrieval studies \citep[e.g.][]{Feng2018, Krissansen-Totton2018, Barstow2020b}. In addition to the ($\log_{10}$) VMRs of \ce{CO2} and \ce{H2O}, we fit for the isothermal temperature $T_0$ (in Kelvin), the solid-body surface reference planet radius $R_0$ (in $R_{\oplus}$), and the surface reference pressure $P_0$ (in $\log_{10}$ Pascals). Although we do not formally include clouds in this study, the reference radius and pressure can be used to account for an opaque gray cloud top. In total, we use a five parameter state vector, $\boldsymbol{\omega}$, for our forward model, subject to the following uninformative priors on $\boldsymbol{\omega}$:  
\begin{equation}
\mathcal{P}(\boldsymbol{\omega}) 
\begin{cases}
    f_{\ce{H2O}} \sim \mathcal{U}(-12,0) ~ \log_{10}(\mathrm{VMR}) \\
    f_{\ce{CO2}} \sim \mathcal{U}(-12,0) ~ \log_{10}(\mathrm{VMR}) \\
    T_0 \sim \mathcal{U}(50,500) ~ \mathrm{K} \\
    R_0 \sim \mathcal{U}(0.95,1.05) ~ R_{\oplus} \\
    P_0 \sim \mathcal{U}(2,6) ~  \log_{10}(\mathrm{Pa})
\end{cases}
\end{equation}
where $\mathcal{U}(\mathrm{lower, upper})$ denotes a uniform distribution with finite probability between the lower and upper bounds. 
We use the function $g(\boldsymbol{\omega})$ to denote the forward model transformation of parameters $\boldsymbol{\omega}$ into wavelength dependent spectroscopic units that can be directly compared to the data. 

\subsubsection{Inverse Model}
\label{sec:methods:inverse}

We use the \dynesty nested sampling code \citep{Speagle2020, Skilling2004} to solve the Bayesian inverse problem for the posterior probability distribution function (PDF) of our forward model parameters given transmission spectrum observations. 
A standard $\chi^2$ log-likelihood function is used to calculate the probability of the transmission spectrum data ($\Delta F = (R_p/R_s)^2$) given the model parameters,  
\begin{equation}
    \ln\mathcal{L} = -\frac{1}{2} \sum_{j=1}^{M} \left ( \frac{\Delta F_j - g(\boldsymbol{\omega})}{\sigma_j} \right )^2, 
\end{equation}
where $\sigma_j$ is uncertainty on the spectrum for the $j$th observed wavelength. Adding the log-likelihood to the logarithm of the aforementioned uninformative priors yields the unnormalized log-posterior that can be sampled with \dynesty.  We run \dynesty with 1000 live points and take model convergence to be achieved when the estimated contribution of the remaining prior volume to the total evidence ($\hat{\mathcal{Z}}$) falls below $\Delta \ln \hat{\mathcal{Z}} < 0.5$ between consecutive iterations. This procedure yields $K$ equally weighted samples from the posterior distribution, where $K \approx 16{,}000$ for our experimental setup, as we will see in Section \ref{sec:results:hbar}.   

\subsection{Hierarchical Modeling Approach}
\label{sec:methods:hbar}

We employ the importance sampling hierarchical model described in \citet{Hogg2010} originally presented for inferring the eccentricity distribution of exoplanets. This model has numerous advantages over a traditional fully coupled hierarchical approach. First, all atmospheric retrievals can be pre-computed, independently, and potentially in parallel, using traditional methods and posterior sampling approaches. This assumes that there are no likelihood covariances between parameters from different planets, but benefits computationally from not being required to sample the corresponding high dimensional spaces. 

Following closely with the derivations presented in \citet{Hogg2010}, for any exoplanetary spectrum $n$, there are $\boldsymbol{\omega}_n$ parameters that we attempt to infer 
\begin{equation}
    \boldsymbol{\omega}_n \equiv [ f_{\ce{H2O}n}, f_{\ce{CO2}n}, T_n, R_n, P_n]
\end{equation}
as defined in Section \ref{sec:methods:nominal}. 

Consider that we have $N$ exoplanets $n$ ($1 \le n \le N$), each of which has $M_n$ transmission spectrum measurements $\Delta F_{nj}$ (or wavelength resolution elements). For every exoplanet $n$, the set of spectroscopic measurements
\begin{equation}
    \mathbf{D}_n \equiv \{ \Delta F_{nj} \}_{j=1}^{M_n}
\end{equation}
is modeled using a radiative transfer forward model. This is given by 
\begin{equation}
    \Delta F_{nj} = \left ( \frac{R_p}{R_s} \right )^2 = g_n(\boldsymbol{\omega}_n) + \mathcal{N}(0, \sigma^2_{nj})
\end{equation}
where the function $g_n(\boldsymbol{\omega}_n)$ is the spectroscopic forward model described previously in Section \ref{sec:methods:forward}, which is parameterized in terms of the five aforementioned dimensions of $\boldsymbol{\omega}_n$, and has an additive noise component drawn from a normal distribution ($\mathcal{N}$) with variance $\sigma^2_{nj}$ (the uncertainty of the $j$th observed wavelength of the $n$th star-planet system). The model of all $N$ planets has $5 \times N$ continuous parameters in the larger list $\{ \boldsymbol{\omega}_n \}_{n=1}^{N}$. We note that it is the high dimensional $5 \times N$ parameter space that makes a fully coupled hierarchical model (e.g. using \texttt{PyMC3}) computationally inefficient, as it must simultaneously infer all $5 \times N$ parameters.  

The likelihood $\mathcal{L}_n$ for the five parameters $\boldsymbol{\omega}_n$ for the $n$th exoplanet spectrum is the probability of the data $\mathbf{D}_n$ for planet $n$ given the parameters $\boldsymbol{\omega}_n$ 
\begin{align}
    \mathcal{L}_n \equiv \mathcal{P}( \mathbf{D}_n | \boldsymbol{\omega}_n). 
\end{align}
This is the standard Bayesian likelihood function previously defined in Section \ref{sec:methods:inverse}. Now for each planet $n$, suppose that we have inferred (using the inverse model described in Section \ref{sec:methods:inverse}) or been provided with a $K$-element sample from a posterior PDF created from the likelihood and an uninformative prior PDF $\mathcal{P}_0(\boldsymbol{\omega}_n)$:
\begin{equation}
\label{eqn:bayes_thm}
    \mathcal{P}(\boldsymbol{\omega}_n | \mathbf{D}_n) = \frac{\mathcal{P}( \mathbf{D}_n | \boldsymbol{\omega}_n) \mathcal{P}_0(\boldsymbol{\omega}_n)}{Z_n}
\end{equation}
where $Z_n$ is the marginal likelihood or evidence, and is simply a normalization constant. We expect that the prior PDF $\mathcal{P}_0(\boldsymbol{\omega}_n)$ is uninformative. For every exoplanet $n$ this posterior sampling is a collection of $K$ equally weighted samples $k$, each itself a set of five parameters $\boldsymbol{\omega}_{nk}$. 
The joint likelihood $\mathcal{L}$ of all parameters for all exoplanets $n$ in the dataset is simply the product of the individual likelihoods: 
\begin{align}
    \mathcal{L} &\equiv \mathcal{P} \left ( \{ \mathbf{D}_n \}_{n=1}^{N} | \{ \boldsymbol{\omega}_n \}_{n=1}^{N} \right )  \\
    &= \prod_{n=1}^{N} \mathcal{L}_n. 
\end{align}
As stated in \citet{Hogg2010}, this makes the assumption that there are no likelihood covariances between the parameters of different exoplanets $n$. 

Now we want to reframe the problem slightly and instead consider the likelihood $ \mathcal{L}_{\alpha}$ for the set of (hyper)parameters $\boldsymbol{\alpha}$ that are used to define an updated prior probability on the \ce{CO2} abundance $\mathcal{P}_{\alpha}(f_{\ce{CO2}})$. It is this updated prior that will be used to encode and characterize population-level trends in atmospheric parameters, which, once known, is a better choice of prior on \ce{CO2} than our original uninformative choice. The probability of the entire data ensemble given the population-level parameters $\boldsymbol{\alpha}$ is then given by: 

\begin{align}
    \mathcal{L}_{\alpha} &\equiv \mathcal{P} \left ( \{ \mathbf{D}_n \}_{n=1}^{N} | \boldsymbol{\alpha} \right ).
\end{align}
This joint likelihood can be expressed as the product of $N$ marginalization integrals over parameters $\boldsymbol{\omega}_n$,  
\begin{align}
    \mathcal{L}_{\alpha} &= \prod_{n=1}^{N} \int \mathcal{P}(\mathbf{D}_n, \boldsymbol{\omega}_n | \boldsymbol{\alpha}) \mathrm{d}\boldsymbol{\omega}_n \\
    &= \prod_{n=1}^{N} \int \mathcal{P}(\mathbf{D}_n |  \boldsymbol{\omega}_n , \boldsymbol{\alpha}) \mathcal{P}(\boldsymbol{\omega}_n | \boldsymbol{\alpha})  \mathrm{d}\boldsymbol{\omega}_n, 
\end{align}
which can be factored further by assuming that the data $\mathbf{D}_n$ depend on $\boldsymbol{\alpha}$ only through $\boldsymbol{\omega}_n$ \footnote{This is a common assumption in hierarchical Bayesian inference, since population-level hyperparameters are typically used to constrain the priors on the physical parameters at the individual level, rather than to directly modify the observables.} (i.e., $\mathcal{P}(\mathbf{D}_n |  \boldsymbol{\omega}_n , \boldsymbol{\alpha}) = \mathcal{P}(\mathbf{D}_n | \boldsymbol{\omega}_n)$), such that 
\begin{align}
\label{eqn:Lalpha_intermediate}
    \mathcal{L}_{\alpha} &= \prod_{n=1}^{N} \int \underbrace{\mathcal{P}(\mathbf{D}_n | \boldsymbol{\omega}_n)}_{\mathcal{L}_n} \mathcal{P}(\boldsymbol{\omega}_n | \boldsymbol{\alpha}) \mathrm{d}\boldsymbol{\omega}_n. 
\end{align}
We recognize the first term in the integrand as the individual likelihood for the $n$th planet spectrum, but the second term---the probability of $\boldsymbol{\omega}_n$ given $\boldsymbol{\alpha}$---is critical and allows us to re-weight the integral using our new hierarchical prior,  
\begin{equation}
\label{eqn:reweight}
    \mathcal{P}(\boldsymbol{\omega}_n | \boldsymbol{\alpha}) \equiv \frac{\mathcal{P}_{\alpha}(f_{\ce{CO2}n}) \mathcal{P}_0(\boldsymbol{\omega}_n)}{\mathcal{P}_0(f_{\ce{CO2}n})}.
\end{equation} 
\Cref{eqn:reweight} divides out the contribution from \ce{CO2} to the original uninformative prior and multiplies through by the new \ce{CO2} prior that depends on the hyperparameters $\boldsymbol{\alpha}$. 
Substituting \Cref{eqn:reweight} into \Cref{eqn:Lalpha_intermediate} and recognizing that the product of the individual likelihood times the original uninformative prior is the posterior via \Cref{eqn:bayes_thm}, we arrive at the following $N$ multidimensional integrals:   
\begin{align}
\label{eqn:hbar_like_int}
    \mathcal{L}_{\alpha} &= \prod_{n=1}^{N} \int \mathcal{P}( \boldsymbol{\omega}_n | \mathbf{D}_n)   \frac{\mathcal{P}_{\alpha}(f_{\ce{CO2}n})}{\mathcal{P}_0(f_{\ce{CO2}n})} \mathrm{d}\boldsymbol{\omega}_n. 
\end{align}
Note that we have dropped the dependence of \Cref{eqn:hbar_like_int} on the marginal likelihood $Z_n$ because typical posterior inference applications only evaluate \Cref{eqn:bayes_thm} up to the unknown normalization constant. 

While \Cref{eqn:hbar_like_int} looks computationally exhausting, the fact that we have already obtained posterior samples simplifies the integral substantially. As articulated in \citet{Hogg2010}, since all probability integrals can be approximated as sums over samples, we can employ the posterior sampling approximation to obtain 
\begin{equation}
\label{eqn:hbar_like}
    \mathcal{L}_{\alpha} \approx \prod_{n=1}^{N} \frac{1}{K} \sum_{k=1}^{K} \frac{\mathcal{P}_{\alpha}(f_{\ce{CO2}nk})}{\mathcal{P}_0(f_{\ce{CO2}nk})}
\end{equation}
where $k$ runs over all posterior samples $K$, and the sum simply contains the ratio of the new prior PDF that we want to infer $\mathcal{P}_{\alpha}(f_{\ce{CO2}nk})$ to the uninformative prior PDF that was used in the original retrieval inference. Although the individual posteriors $\mathcal{P}( \boldsymbol{\omega}_n | \mathbf{D}_n)$ do not explicitly appear in \Cref{eqn:hbar_like}, they are implicitly contained within the distribution of $K$ samples. 
With the likelihood $\mathcal{L}_{\alpha}$ as defined in \Cref{eqn:hbar_like}, it is straightforward to infer posteriors on the population-level parameters $\boldsymbol{\alpha}$ using Bayes' Theorem, 
\begin{equation}
\label{eqn:hbar_bayes}
    \mathcal{P} \left ( \boldsymbol{\alpha} | \{ \mathbf{D}_n \}_{n=1}^{N} \right ) \propto \mathcal{L}_{\alpha} \mathcal{P}(\boldsymbol{\alpha}), 
\end{equation}
where $\mathcal{P}(\boldsymbol{\alpha})$ is the (hyper)prior PDF for the hyperparameters $\boldsymbol{\alpha}$.  

Within this section we have kept the importance sampling derivations as agnostic as possible to the specifics of the population-level trend(s) under consideration to ensure that the methods can be readily adapted to other problems. Critically, we have not yet specified the form of the \ce{CO2} trend, the hyperparameters that define $\boldsymbol{\alpha}$, or their respective hyperpriors. We refer the reader to Section \ref{sec:results} (particularly \Cref{eqn:analytic}, \Cref{eqn:hbar_new_prior}, and \Cref{eqn:hyper_prior}) for our specific implementation and the subsequent results. 

In the original \citet{Hogg2010} formulation of importance sampling, it was assumed that the original exoplanet data was not in-hand, but that the posteriors had been obtained from a colleague or another research group for further analysis. While this is certainly a circumstance that may motivate the use of importance sampling for atmospheric retrievals, we found that it was crucial, perhaps necessary, to perform the hierarchical analysis in a subsequent step following the completion of a uniform set of retrievals, due to the excessive computational expense of simultaneously inferring the population parameters $\boldsymbol{\alpha}$ along with all of the $5 \times N$ individual system atmospheric parameters $\boldsymbol{\omega}_n$. This intractability stems from the inherent computational expense of retrieval codes, which solve the radiative transfer equation at each step in the spectral inference. However, retrieval codes with exceptionally fast forward models may prove important for exploring the cost-benefit analysis of HBAR methods that simultaneously infer individual and population parameters. This may be an opportunity for machine learning augmented retrievals \citep[e.g.,][]{Zingales2018, Nixon2020, Himes2020, Hayes2020}. 

\section{Results} 
\label{sec:results}

First, in Section \ref{sec:results:spectra}, we present a uniform set of spectral models for an idealized population of terrestrial exoplanets that by design exhibit the silicate weathering \ce{CO2} trend of interest. Second, in Section \ref{sec:results:hbar}, we use our new HBAR model to infer the atmospheric \ce{CO2} trend across the population of synthetic exoplanets. Third, in Section \ref{sec:results:compare}, we conduct a population-level model comparison to determine the robustness of the inferred trend relative to the null hypothesis and other functional forms for the silicate weathering relation.  

\subsection{Spectral Models}
\label{sec:results:spectra}

We generate a set of transmission spectra that will allow us to empirically test for the existence of the habitable zone as described in \cite{Bean2017}. To limit the number of confounding factors in this study, we assume that the set of $N$ exoplanets with observed spectra are identical to one another except for their stellar irradiation and the quantity of \ce{CO2} and \ce{N2} in their atmospheres, which follows the silicate weathering feedback trend that we impose. We assume the planets possess 1 bar Earth-like atmospheres composed only of \ce{N2}, \ce{CO2}, and \ce{H2O}. We use globally averaged Earth vertical thermal and \ce{H2O} profiles \citep[from][]{Robinson2010, Robinson2011, Schwieterman2015} to satisfy the assumption that each planet is habitable, but neglect changes in these profiles that would be expected from self-consistent climate modeling across the HZ to limit our focus to observables due to \ce{CO2}. Furthermore, we assume that all simulated planets 
%share physical characteristics (mass, radius, transit depth and duration) with the current state of knowledge on 
are Earth-sized (1 M$_{\oplus}$, 1 R$_{\oplus}$) and orbit TRAPPIST-1 with the same transit duration as 
TRAPPIST-1e \citep{Gillon2017, Agol2021}, which simply provides a tangible point of comparison to judge the plausibility of such an analysis in the context of current exoplanet targets and observing capabilities.  

To simplify the underlying model for our population trend, we fit an analytic function to the predicted \ce{CO2} volume mixing ratios calculated by \citet{Bean2017} using a 1D radiative-convective climate model. We used the following ``Gaussian-like'' functional form 
\begin{equation}
\label{eqn:analytic}
    f_{\ce{CO2}}(\mu, \sigma_1, \delta) = \frac{1}{\delta \sqrt{2 \pi}} \exp \left [ -0.5 \left ( \frac{S_{\oplus} - \mu}{\sigma_1} \right )^{4} \right ]
\end{equation}
where $S_{\oplus}$ is the stellar irradiation incident on the planet relative to Earth and $\mu$, $\sigma_1$, and $\delta$ are free parameters which we determine to be 0.04727, 0.5372, and 0.4376 respectively by minimizing the squared residuals. We assume that the remainder of the atmospheric volume is filled with \ce{N2}, and then calculate the mean molecular weight of the atmosphere self-consistently. 

\begin{figure*}[t!]
\centering
\includegraphics[width=1.0\textwidth]{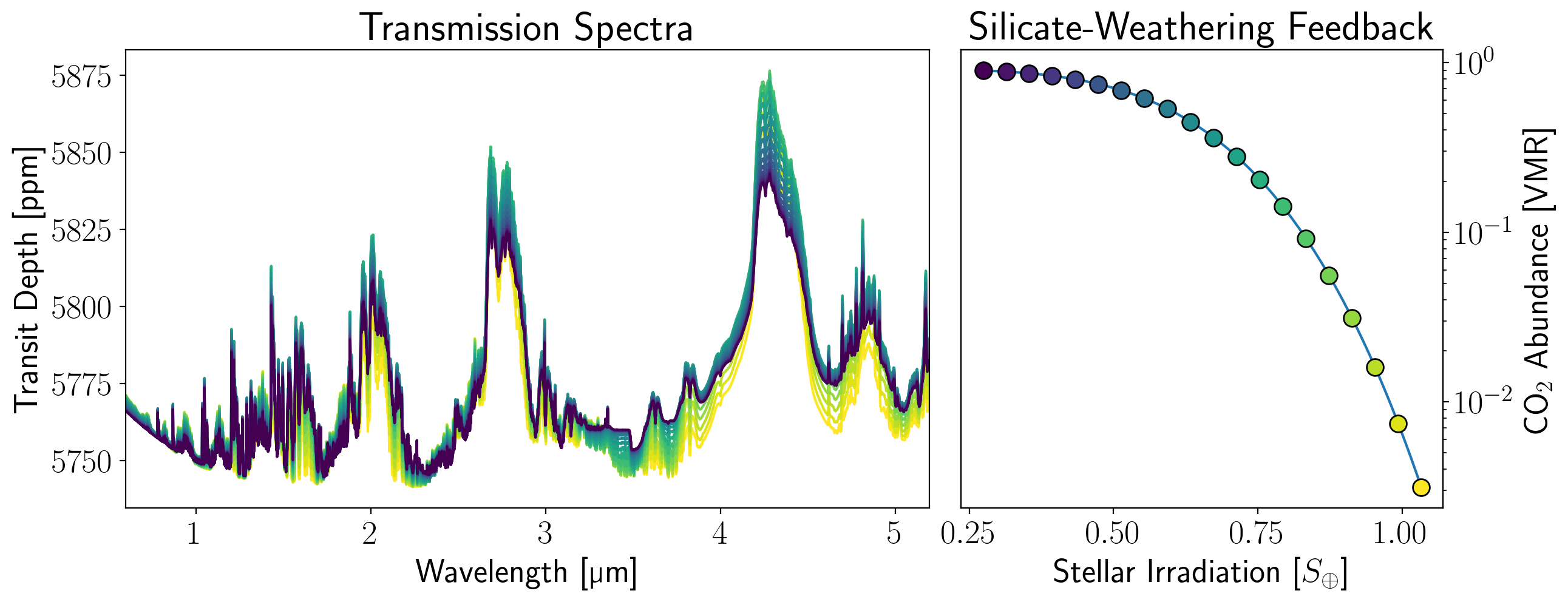}
\caption{Transmission spectrum models (left) for a sample of 20 hypothetical rocky exoplanets generated following trend in \ce{CO2} with stellar irradiation predicated on the assumption of a functioning carbonate-silicate weathering feedback mechanism (right). Planets near the outer edge of the HZ at lower stellar irradiation are predicted to possess atmospheres with higher \ce{CO2} abundances than those near the inner edge of the HZ, and such variations in \ce{CO2} manifest in observable features in the transmission spectrum.}
\label{fig:weathering_spectra}
\end{figure*}

\Cref{fig:weathering_spectra} shows our resulting transmission spectrum models at 1 cm$^{-1}$ wavenumber resolution (left panel) that correspond to the assumed trend in \ce{CO2} with stellar irradiation (right panel) due to the carbonate-silicate weathering feedback mechanism for $N=20$ theoretical exoplanets. By design, the spectra exhibit differences due solely to the volume mixing ratio of \ce{CO2} (and implicitly \ce{N2}), which are small relative to the total transit depth. Two competing effects shape the observable characteristics of the spectra shown in \Cref{fig:weathering_spectra}: the \ce{CO2} optical depth and the atmospheric mean molecular weight. As the \ce{CO2} abundance increases the \ce{CO2} optical depth increases, and the weak \ce{CO2} bands, primarily seen between $1 - 2$ \um, increase in absorption strength. The opposite is seen for the saturated \ce{CO2} bands at 2.7 and 4.3 \um. The increase in \ce{CO2} causes the saturated bands to decrease in absorption strength as the mean molecular weight of the atmosphere increases from \ce{N2}-dominated (28 g/mol) to \ce{CO2}-dominated (44 g/mol), and the atmospheric scale height decreases correspondingly. In general, these subtle spectral differences must be sufficiently resolved in each observed spectrum for the population-level model to infer a meaningful trend. 

\begin{figure}[t]
\centering
\includegraphics[width=0.47\textwidth]{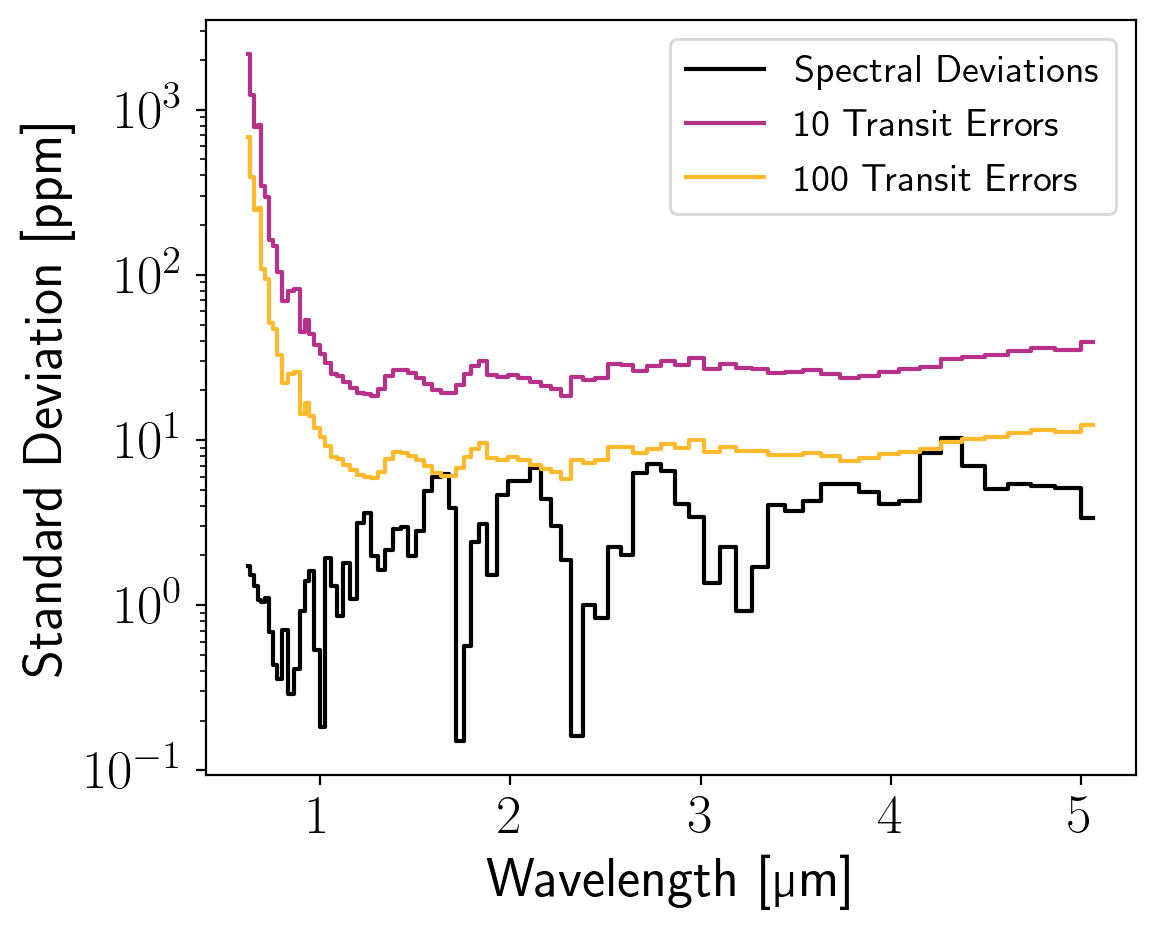}
\caption{Standard deviation among the spectra shown in \Cref{fig:weathering_spectra} with varying \ce{CO2} (black) compared with the $1 \sigma$ observational uncertainties for 10 (magenta) and 100 (orange) stacked transits of TRAPPIST-1e with JWST's NIRSpec Prism. The spectroscopic differences caused by the silicate weathering \ce{CO2} trend are less than 10 ppm for Earth-sized exoplanets transiting TRAPPIST-1, indicating that approximately 100 stacked transits may be necessary to clearly resolve the deviations in \ce{CO2} mixing ratios across HZ instellations.}
\label{fig:precision}
\end{figure}

We used the \pandexo JWST noise model \citep{Batalha2017b, Pandexo2018} to simulate synthetic transmission spectrum observations using the Near-Infrared Spectrograph (NIRSpec) Prism instrument \citep{Bagnasco2007, Ferruit2014}. We used the same \pandexo simulation setup as \citet{Lustig-Yaeger2019} assuming the partial saturation strategy for the NIRSpec Prism \citep{Batalha2018} and no assumed noise floor. \Cref{fig:precision} shows the precision of the Prism spectra for TRAPPIST-1e using 10 and 100 stacked transits, compared against the standard deviation among the spectra shown in \Cref{fig:weathering_spectra}. Based on the fact that the $1\sigma$ spectral uncertainties for 100 transits is of similar magnitude to the deviations caused by \ce{CO2}, we conclude that approximately 100 observed transits may be required, for each planet, to obtain spectra with high enough precision to clearly resolve the \ce{CO2} bands in sufficient detail to distinguish between the atmospheres and resolve the trend. While this is an objectively large number even for a single target, we adopt the spectral uncertainties corresponding to 100 stacked transits for each target to ensure that the next stage in the analysis will contain enough information to properly test this population trend with our HBAR model.  

\subsection{An Empirical Test of the Habitable Zone} 
\label{sec:results:hbar}

\begin{figure*}[t]
\centering
\includegraphics[width=0.97\textwidth]{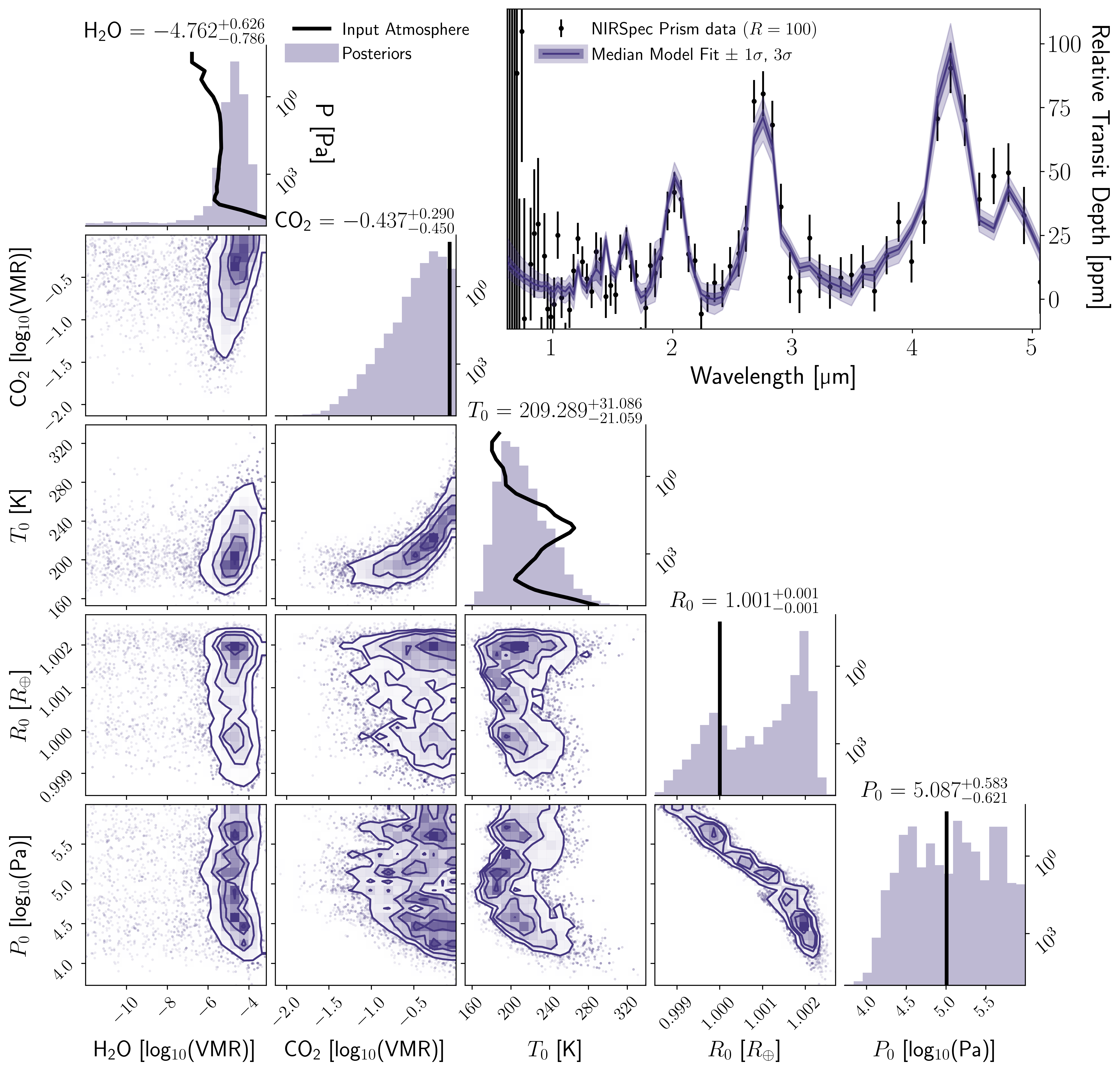}
\caption{Corner plot showing the 1D and 2D marginalized posterior distributions (histograms and contours) for the five planetary parameters retrieved from fitting the transmission spectrum using the \smarter model. The subplots along the diagonal show the input atmospheric profiles as a function of pressure that were used to generate the synthetic spectrum for comparison against the vertically homogeneous (isothermal and evenly mixed) atmospheric models that were retrieved. The upper right inset shows the median retrieved spectrum bounded by $1 \sigma$ and $3 \sigma$ credible intervals derived from posterior samples. This particular retrieval result was for the case with ${\sim}83\%$ \ce{CO2} and is representative of similar results obtained for the other planets in our synthetic sample.}
\label{fig:retrieval_corner}
\end{figure*}

We performed a uniform set of retrievals on the 20 spectra shown in \Cref{fig:weathering_spectra} that evenly span the habitable zone range of stellar irradiation with spectral resolution and noise appropriate for 100 transits with JWST/NIRSpec Prism. \Cref{fig:retrieval_corner} shows a representative corner plot of the 1D and 2D marginalized posterior distributions for the inferred physical planetary parameters for the case with the ${\sim}83\%$ atmospheric \ce{CO2}---the fourth largest \ce{CO2} abundance in the sample. The covariance between the isothermal temperature and \ce{CO2} abundance shows a degeneracy due to the dependence of the atmospheric scale height on the temperature and mean molecular weight. Similarly the reference radius and pressure show an expected degeneracy that represents the set of radii and pressures that maintain the transmission spectrum continuum near the observed value. The retrieved \ce{H2O} abundance is consistent with stratospheric values and, notably for the purpose of this investigation, the \ce{CO2} abundance is constrained to within $\pm 0.5$ dex. The upper right of \Cref{fig:retrieval_corner} shows the median model transmission spectrum obtained from fitting the synthetic JWST data with bounding envelopes to represent the upper and lower $1 \sigma$ and $3 \sigma$ credible intervals. The spectral models shown were derived from 500 random samples from the posterior distribution.  

Although the retrieval results shown in \Cref{fig:retrieval_corner} are only for one representative case from our sample, the other 19 retrievals show similar results with expected differences caused by the different underlying \ce{CO2} abundance and the propagation of random Gaussian scatter in the spectrum through the inference procedure. On average each retrieval with \dynesty yielded $K \approx 16{,}000$ equally weighted samples from the respective posteriors. Next, the ensemble of posteriors obtained from the individual planets will be used to infer population-level parameters, in an attempt to retrieve the silicate weathering feedback trend that we injected into the sample.   

With the posteriors from our uniform retrieval analysis in hand, we now wish to infer population-level parameters for the \ce{CO2} versus stellar irradiation trend by running MCMC on the HBAR importance sampling model. The HBAR likelihood function is given by \Cref{eqn:hbar_like}, where the original prior $\mathcal{P}_0(f_{\ce{CO2}nk})$ on the $\log$\ce{CO2} abundance is uninformative $\mathcal{U}(-12,0)$ and the updated prior $\mathcal{P}_{\alpha}(f_{\ce{CO2}nk})$ is calculated from the analytic relationship provided in \Cref{eqn:analytic}. Specifically, the updated prior is a function of hyperparameters $\boldsymbol{\alpha}$ and is taken to be normally distributed, 
\begin{equation}
\label{eqn:hbar_new_prior}
    \mathcal{P}_{\alpha}(f_{\ce{CO2}nk}) = \mathcal{N}(f_{\ce{CO2}nk} - f_{\ce{CO2}}(\mu, \sigma_1, \delta), \sigma_2),
\end{equation}
where $\sigma_2$ is the standard deviation of the Gaussian distribution that lends high probability to values of the \ce{CO2} population trend that lie closest to the original \ce{CO2} posterior samples. Thus, for our HBAR importance sampling model, we have the free hyperparameters $\boldsymbol{\alpha} \equiv [\mu, \sigma_1, \delta, \sigma_2]$ that we seek to infer, subject to the following uninformative hyperpriors on $\boldsymbol{\alpha}$: 
\begin{align}
\label{eqn:hyper_prior}
\mathcal{P}(\boldsymbol{\alpha})
\begin{cases}
    \mu \sim \mathcal{U}(-2,2) \\
    \sigma_1 \sim \mathcal{U}(0,2) \\
    \delta \sim \mathcal{U}(0,2) \\
    \sigma_2 \sim \mathcal{N}_{1/2}(0,1),
\end{cases}
\end{align}
where $\mathcal{N}_{1/2}(0,1)$ refers to a half-normal distribution with a mean of 0 and a standard deviation of 1. At this point, we are ready to evaluate our HBAR model and infer the hyperparameters $\boldsymbol{\alpha}$. To recap, we now have a specific population-level model for the \ce{CO2} trend (\Cref{eqn:analytic}) that is used to define a new population-level prior on the \ce{CO2} abundance (\Cref{eqn:hbar_new_prior}). This allows the likelihood $\mathcal{L}_{\alpha}$ to be obtained by evaluating \Cref{eqn:hbar_like}. Finally, the uninformative hyperprior (\Cref{eqn:hyper_prior}) can be multiplied by the likelihood, as in \Cref{eqn:hbar_bayes}, to infer the desired $\boldsymbol{\alpha}$ posterior PDF. 

We used MCMC with \emcee \citep{Foreman-Mackey2013} to infer posterior samples of the hyperparameters $\boldsymbol{\alpha}$ using 20 walkers. We ran the chain until it reached a length of approximately $50 \times$ the integrated autocorrelation time, as suggested by the \emcee code documentation\footnote{\url{https://emcee.readthedocs.io/en/stable/tutorials/autocorr/}}. 

\begin{figure*}[t]
\centering
\includegraphics[width=0.97\textwidth]{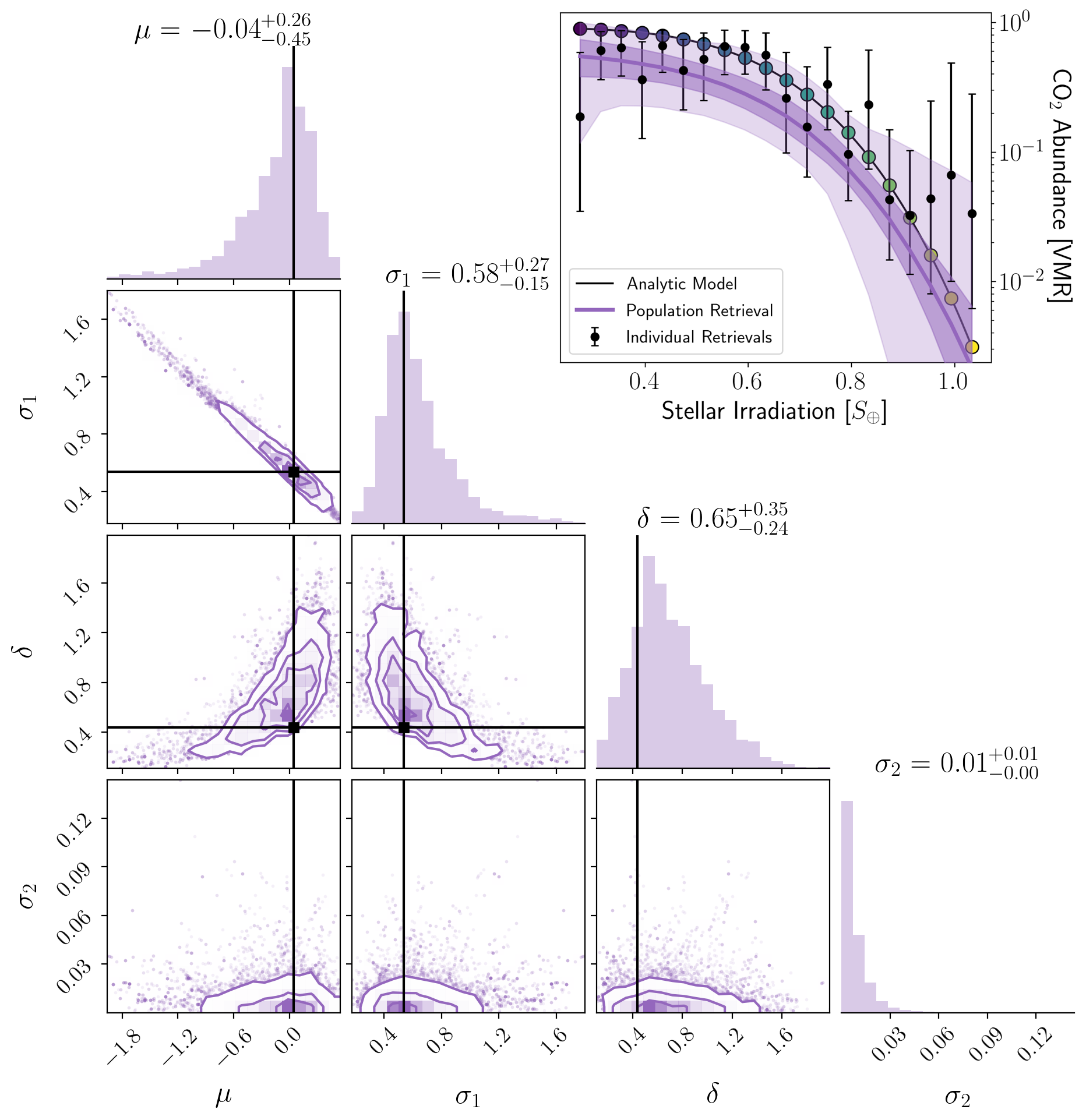}
\caption{Corner plot showing 1D and 2D marginalized posterior distributions for the four hyperparameters used to describe the carbonate-silicate weathering feedback population-level trend for rocky HZ exoplanets. The median inferred population trend is displayed in the upper right inset (solid purple line) with shading to denote $1 \sigma$ and $3 \sigma$ credible intervals. Using the importance sampling HBAR method we are able to retrieve statistically robust constraints on the parameters characterizing the silicate weathering population trend in \ce{CO2} abundance with stellar irradiation.}
\label{fig:weathering_hbar_corner}
\end{figure*}

\Cref{fig:weathering_hbar_corner} shows the MCMC results from our HBAR importance sampling model. 
The lower left set of panels in \Cref{fig:weathering_hbar_corner} show the 1D and 2D marginalized posteriors for the population-level hyperparameters in our HBAR model. Relative to their uninformative hyperpriors, the hyperparameters are well constrained by the inference. The median retrieved trend in \ce{CO2} with stellar irradiation is calculated from the posterior samples and is shown in the upper right panel of \Cref{fig:weathering_hbar_corner}, bounded by the 1$\sigma$ and 3$\sigma$ credible intervals. 
The upper right panel also shows the retrieved $1 \sigma$ \ce{CO2} constraints for all 20 independent spectrum retrievals plotted as a function of stellar irradiation. It may be conceptually useful to imagine that we have directly fit the purple population model to the black error bars in \ce{CO2} abundance, while in fact we have actually taken the full set of multidimensional posteriors into consideration in our numerical evaluation of \Cref{eqn:hbar_like} and \Cref{eqn:hbar_bayes}. The true underlying \ce{CO2} trend is also shown for reference, where the characteristic decline in \ce{CO2} abundance with stellar irradiation is well resolved by the population-level inference. 

\subsection{Model Comparison}
\label{sec:results:compare}

One of the benefits of performing an importance sampling HBAR meta-analysis is that multiple different hypothesized population-level atmospheric trends can be investigated and compared without the need to re-run the computationally expensive retrieval models. We now compare three simpler population models to our previously obtained result to demonstrate how such models can be discriminated. These population-level models for the \ce{CO2} versus stellar irradiation relation include a linear trend, a log-linear trend, and a flat non-trend representing the null hypothesis.  

\begin{figure}[t]
\centering
\includegraphics[width=0.47\textwidth]{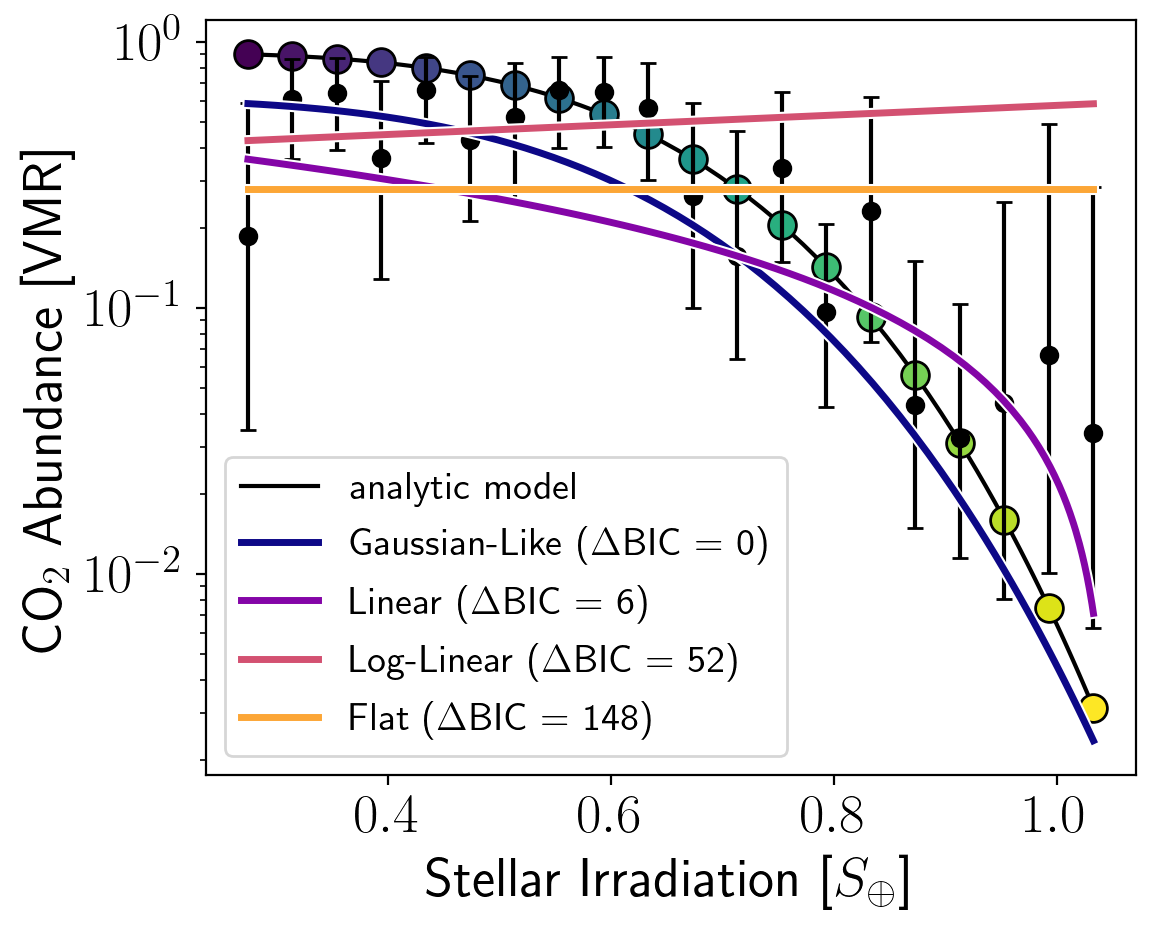}
\caption{Comparison between multiple population-level models. The $\Delta$BIC assessment indicates that the ``Gaussian-like'' model is strongly preferred over the other models.}
\label{fig:model_comparison}
\end{figure}

\Cref{fig:model_comparison} compares four different best-fitting models to the population-level \ce{CO2} trend. The Bayesian Information Criterion (BIC) is calculated for each model using the following relation 
\begin{equation}
    \mathrm{BIC} = \kappa \ln N - 2 \ln \mathcal{\hat{L}_{\alpha}}, 
\end{equation}
where $\kappa$ is the number of free population-level parameters estimated by the model (i.e., the number of dimensions in $\boldsymbol{\alpha}$), $N$ is taken to be the number of individual planet spectra in the population, and $\ln \mathcal{\hat{L}_{\alpha}}$ is maximum of the log-likelihood function obtained through optimization. When selecting between multiple models, the model with the lowest BIC is taken to be preferred. We subtract the preferred model BIC from every other model to obtain $\Delta$BIC values for comparison. The $\Delta$BIC values in \Cref{fig:model_comparison} indicate that the ``Gaussian-like'' model is preferred and there is strong evidence against all of the models with higher BICs. This is the expected result because we generated the synthetic planetary models using the Gaussian-like population trend, but this serves as a useful demonstration of how population-level trends can be compared using an importance sampling HBAR framework.   

\section{Discussion} 
\label{sec:discussion}

We conducted an idealized simulated search for a trend in \ce{CO2} abundance with stellar irradiation predicted by the carbonate-silicate weathering feedback for exoplanets in the HZ. Our approach used a novel hierarchical Bayesian model---a first of its kind for exoplanet atmospheric retrievals---to infer population trends in atmospheric characteristics that may prove useful well beyond the scope of this work. We elaborate on our scientific and methodological findings in the following subsections. 

\subsection{Practical Challenges to an Empirical Test of Carbonate-Silicate Weathering}

We found that the \ce{CO2} trend predicted by carbonate-silicate weathering will be challenging to infer, potentially limiting the observational feasibility of this statistical comparative planetology test of the HZ. As presented in \citet{Bean2017}, the use of a large statistical sample of exoplanets is enabled by making relatively low precision spectroscopic observations of each planet. However, as we have shown, the changes to the transmission spectrum of an Earth-like planet (which this investigation must detect) caused by changes in \ce{CO2} abundance (which are consistent with the predicted trend) are quite small at $< 10$ ppm for TRAPPIST-1e-like planets. As a result, to infer the silicate weathering trend at high confidence, we found that precise transmission spectra were required that corresponded to 100 stacked transit observations with JWST \textit{of each planet} in our idealized 20 planet sample, despite the optimistic assumption that each planet was TRAPPIST-1e-like (while only one such planet is known to exist). Acquiring such a sample of high-precision terrestrial exoplanet transmission spectra is not feasible in JWST's nominal mission lifetime, nor would it be made appreciably easier using a successor to JWST, such as the Origins Space Telescope (OST) concept \citep{OST2019}, due to the high probability that yet-to-be-discovered transiting HZ rocky exoplanets will be found in systems less amenable for atmospheric characterization than those transiting TRAPPIST-1 \citep{Gillon2020}.     

However, our results do not rule out the possibility of using JWST quality transmission spectra to detect an increase in \ce{CO2} abundance with decreasing insolation. Using our novel HBAR framework, we demonstrated that an optimistic simulated survey would be capable of ruling out the null hypothesis for the silicate weathering feedback \ce{CO2} trend at high confidence. Thus, it stands to reason that fewer transits per planet would be able to resolve the \ce{CO2} population trend with less confidence, while remaining statistically robust. While our results do not reveal the spectral precision of such a transition in statistical confidence, hierarchical Bayesian methods are well-suited to resolve population trends that are not as vividly resolved at the individual level, in particular, for noisy datasets where priors dominate the inference. 

The difficulty of precisely measuring \ce{CO2} \textit{abundances} starkly contrasts against the relative ease with which \ce{CO2} \textit{detections} are predicted for terrestrial exoplanet transmission spectra with JWST. Numerous reports suggest that \ce{CO2} may be an optimal molecule to target to detect the presence of rocky exoplanet atmospheres \citep{Meadows2018, Lustig-Yaeger2019, Fauchez2019, Pidhorodetska2020}. These proposed atmospheric detections hinge upon the strong and saturated \ce{CO2} bands at 4.3 \um and 15 \um, which are largely insensitive to significant changes in \ce{CO2} abundance \citep{Barstow2016, Wunderlich2020}. This places a limit on the \ce{CO2} abundance precision that can be retrieved from the spectrum. One method for overcoming this limitation and inferring more precise \ce{CO2} abundances is to resolve and detect (or confidently non-detect) the weaker \ce{CO2} bands in the NIR that are not saturated using precise transmission spectra, as we have shown here.   

While we have focused exclusively on transmission spectroscopy, other spectroscopic methods for exoplanet atmospheric characterization may prove more successful at detecting population trends in \ce{CO2} abundance. For example, a next-generation direct-imaging mission that can obtain spectra of Earth-like exoplanets around Sun-like stars \citep[as recommended by the][]{Astro2020}, such as the Large UV/Optical/IR Surveyor \citep[LUVOIR;][]{LUVOIR2019} or the Habitable Exoplanet Observatory \citep[HabEx;][]{HabEx2019} with access to the 1.6 \um and 2 \um \ce{CO2} bands, may offer more leverage for precise \ce{CO2} abundance retrievals. However, this will need to be demonstrated in a future study since these weak \ce{CO2} bands were omitted from the seminal retrieval work of \citet[][]{Feng2018} due to the relative insignificance of \ce{CO2} in the Earth's visible and NIR spectrum.  

Nature is likely to produce more complicated atmospheric trends than what we have investigated here. This may hold particularly true for habitable exoplanets where the presence of stable surface liquid water is appreciated to be dependent on a complex interplay of stellar, planetary, and planetary system-wide factors that may indeed produce an elusive population trend that spans many dimensions \citep{Meadows2018c}. To this end, recent work by \citet{Lehmer2020} demonstrated that the carbonate-silicate weathering feedback trend in \ce{CO2} with incident flux may be log-linear in form with significant scatter due to individual planet considerations, such as land area for weathering and \ce{CO2} outgassing fluxes. The log-linear trend differs from the non-linear trend from \citet{Bean2017} considered in this work because the \citet{Lehmer2020} model included temperature and \ce{CO2} feedbacks that cause the surface temperature to decline with semimajor axis \citep{Kadoya2014}, rather than remain fixed at 289 K throughout the HZ \citep{Bean2017}. The exact nature of the predicted \ce{CO2}-flux trend in the HZ is unlikely to change our results due to the relative consistency of the two similar hypotheses compared to our posterior constraints on \ce{CO2}. Moreover, these predictions are all likely to be incorrect at some level due to their exclusive reliance on geophysical evidence from Earth. This only further motivates the need for methods that allow us to update our understanding of comparative planetology trends using exoplanet data. Future work could leverage the HBAR framework presented here to investigate the 2D population density trend suggested by \citet{Lehmer2020} and potentially incorporate a third dimension for the surface temperature to better capture the climatic feedbacks expected throughout the HZ \citep{Kadoya2014, Lehmer2020}. 

Similarly, \citet{Seales2021} used coupled geophysical models to study the temporal onset of habitability and found that variations in tectonic efficiency from one planet to another may produce a predictable distribution in the \ce{CO2} abundance for an ensemble of planets with the same absolute age. Thus it may be possible to expand the population trends studied here into the system age dimension to relate an inferred distribution in \ce{CO2} back to predictions from geophysical models. 

\subsection{The Future of HBAR}

We have taken a first step towards developing a hierarchical Bayesian atmospheric retrieval model for tracking population trends in exoplanet atmospheres through the complicated, non-linear, and degenerate problem of fitting exoplanet spectra. Standard atmospheric retrievals are well known to be computationally expensive due to the requirement of a radiative transfer forward model to fit spectroscopic observations. By using the relatively simple importance sampling method for hierarchical Bayesian modeling from \citet{Hogg2010}, we effectively avoid the computationally taxing need to perform retrievals on each planet's spectrum simultaneously, as would be required for a standard hierarchical model. Instead, importance sampling allows for population-level inferences in a straightforward meta-analysis of the posterior samples obtained from a uniform set of standard atmospheric retrieval results. This effectively decouples the computationally expensive retrieval modeling from the hierarchical modeling, such that the hierarchical problem can be readily solved once a uniform set of retrieval results (posteriors) are in-hand. Thus, intrigued readers may find that they already have all of the necessary ingredients to characterize population-level trends within their existing retrieval results. 

Characterizing population-level trends in exoplanet atmospheres has use-cases well beyond the habitable zone and offers a critical capability for advancing comparative planetology with current, upcoming, and future telescopes. For example, population studies of extrasolar gas giants are already underway. Seminal work by \citet{Sing2016} analyzed the spectra of 10 hot Jupiters observed with HST and found that the planets exhibited a continuum from clear to cloudy atmospheres which may suggest that clouds and hazes, rather than water depletion during formation, are the cause of weaker-than-expected \ce{H2O} absorption features. However, subsequent uniform retrieval analyses by \citet{Barstow2017} and \citet{Pinhas2019} complicate this picture as their retrieved \ce{H2O} abundances suggest subsolar oxygen and/or supersolar C/O ratios with no clear correlations identified. Additionally, \citet{Tsiaras2018} conducted a population study of 30 gaseous exoplanets and found that about half of the sample had detectable atmospheres via \ce{H2O} absorption features. Future work on this hot Jupiter sample may benefit from the HBAR model described here to infer and compare population trends for different proposed formation and evolutionary pathways. 
Moving towards smaller planets, \citet{Changeat2020} conducted a uniform retrieval analysis to demonstrate that the ESA-Ariel mission \citep{Tinetti2016} will be sensitive to trends between the atmospheric chemistry and planetary parameters for a population consisting of mostly sub-Neptune and Neptune size planets \citep{Edwards2019}. As more telescopes dedicated to exoplanet atmospheric characterization come online, and as the number of observed exoplanet spectra grows, the use of HBAR modeling may become crucial for comparative planetology.

\section{Conclusion} 
\label{sec:conclusion}

We implemented a first-of-a-kind hierarchical Bayesian atmospheric retrieval model to characterize population-level trends in exoplanet atmospheres.  
We argue that hierarchical Bayesian models are well suited for this task due to the sophisticated inference methods (retrievals) required to transform the observed spectra of exoplanets into meaningful atmospheric characteristics. In particular, the HBAR model that we implemented using importance sampling offers a computationally tractable approach for performing such multi-level inferences because it requires only the posteriors from a uniform set of traditional retrievals, which can all be performed independently. Marginalizing over the full multidimensional posteriors allows the importance sampling HBAR method to propagate complicated parameter covariances through to the population-level hyperparameters; conserving information that may be lost when analyzing retrieved atmospheric trends using only 1D marginalized posteriors or traditional statistical moments. While this study by no mean represents the end-all-be-all of HBAR modeling, we have taken the first few steps towards a computationally tractable HBAR model that will benefit over time from further application and refinement by the exoplanet community.

We tested the importance sampling HBAR framework on an empirical probe of the HZ using simulated transmission spectra of rocky planets with an injected trend in \ce{CO2} abundance with stellar irradiation that is consistent with predictions for a functioning carbonate-silicate weathering negative feedback cycle. We demonstrated that the HBAR method can be used to (1) accurately constrain population-level parameters that characterize the silicate weathering trend and (2) discriminate between multiple different hypothetical population trends. However, we found that such precise spectroscopic measurements would be required to sense the \ce{CO2} trend in terrestrial exoplanet atmospheres that inferring this particular statistical comparative planetology trend may be infeasible using upcoming missions with transmission spectroscopy capabilities. Nonetheless, the use of the HBAR methods presented here may prove to be an important ingredient for future comparative planetology studies as new theories of planetary atmospheric formation, evolution, and habitability are forged in the crucible of exoplanet demographics. 

\acknowledgments

We would like to thank D. Foreman-Mackey for illuminating numerous dense statistical concepts, C. Tinsman for supporting our use of APL's computing resources, and the anonymous reviewer for helping us improve the quality of this manuscript. This work was funded by internal research and development funding from the Johns Hopkins Applied Physics Laboratory. 

\software{Astropy \citep{Astropy2013, Astropy2018}, corner \citep{corner}, Dynesty \citep{Speagle2020}, emcee \citep{Foreman-Mackey2013}, LBLABC \citep{Meadows1996}, Matplotlib \citep{Hunter2007}, NumPy \citep{Walt2011}, SciPy \citep{Virtanen2019scipy}, SMART \citep{Meadows1996}, Pandas \citep{pandas2010}, Pandeia \citep{Pontoppidan2016}, PandExo \citep{Batalha2017b, Pandexo2018}, pysynphot \citep{STScI2013}}

\bibliography{ms}

\end{document}